\documentstyle[11pt]{article}

\def\etal{{\underline{et al.} }}
\def\ros{{\sl ROSAT }}
\def\exo{{\sl EXOSAT }}
\def\ein{{\sl Einstein }}
\def\grad{$^\circ$}
\def\Ni{\noindent}
\def\amin{\ifmmode ^{\prime}\else$^{\prime}$\fi}
\def\fdg{\hbox{$.\!\!^\circ$}}             

\begin{document}

\unitlength1mm

\centerline{\huge\bf Search for GRB counterparts}

\vspace{39pt}

\centerline{J. GREINER}

\vspace{26pt}

\centerline{\it Max-Planck-Institute for Extraterrestrial Physics, 85740
Garching, Germany}

\vspace{39pt}

\centerline{\bf INTRODUCTION -- SEARCH STRATEGIES}

\vspace{13pt}

Historically, the search for $\gamma$-ray burst (GRB) counterparts began with
the check of photographic exposures taken contemporaneously with the
burst event$^{1}$.
Due to the lack of obvious associations of the GRB positions with well-known
transients such as supernovae or flare stars, the search for correlations
rapidly spread to very different energy ranges and detection methods.
One example is the search for absorption events in the VLF (164 kHz)
propagation
in the Earth ionosphere caused by the transit of (ionising) X-ray emission
from the bursts$^{2}$, which earlier proved successful for the detection
of Sco X-1.

After these early (mid-seventies) attempts of simultaneous observations
later on (eighties) the
counterpart search was dominated by searches for non-correlated flaring as
well as quiescent sources in different wavelength bands. This kind of
investigation always used the implicit assumption that bursts repeat
on timescales less than 50 years.
Searches for  archival optical transients in small GRB error boxes using
large photographic plate collections were initiated at Harvard
Observatory$^{3}$
and then also performed at other observatories$^{4,5}$. In the same manner,
data of the \ein, \exo and \ros satellites were used to search for
quiescent soft X-ray sources$^{6-10}$.
In parallel, very few of the well-localised
GRB error boxes of the second interplanetary network
were examined with larger telescopes to search for faint
quiescent optical counterparts$^{11,12}$. No reliable counterpart candidate
emerged from all these studies.

With the launch of the {\it Compton} Gamma-Ray Observatory (GRO) and
the BATSE (Burst and Transient Source Experiment) measurements of about
one burst per day improved possibilities emerged for the
correlation with other (simultaneously performed) observing programmes.
This includes the correlation of GRBs with
(1) regularly exposed wide-field photographs of most major
observatories$^{13}$,
(2) a CCD-based multi-camera monitoring of 0.75 ster of the sky$^{14,15}$,
(3) atmospheric airshower events due to TeV emission $^{16-19}$,
(4) neutrino or muon detection rates$^{20-22}$.
Due to an unexpected incident (the failure of the GRO tape
recorders) and the advent of fast data transmission, rapid follow-up
observations of GRB error boxes became possible with the development
of BACODINE$^{23}$ and the COMPTEL/BATSE/NMSU network$^{24}$.
In the following I will concentrate on the new developments connected with
these rapid follow-up observations as well as
the availability of times and locations of many bursts for correlated
investigations (see also other recent reviews$^{25,26})$.

\vspace{26pt}

\centerline{\bf SIMULTANEOUS OBSERVATIONS}

\vspace{13pt}

\centerline{\it\bf Monitoring with Dedicated Automated Telescopes}

\vspace{8pt}

The Explosive Transient Camera (ETC) is a wide (0.75 ster) field-of-view (FOV)
CCD array consisting of 8 camera pairs which operates entirely by computer
control on Kitt Peak National Observatory$^{27}$.
During its more than 4 years
of operation there were five cases in which a BATSE GRB occurred during
an ETC observation and within or near an ETC FOV$^{14,15}$.
No optical transients were detected during these observations, resulting in
upper limits for the {\bf fluence}
ratio L$_\gamma$/L$_{opt}$ $\ge$ 2--120$^{14,15}$.
Unfortunately, in all cases only a part (20\%--80\%) of the rather large
BATSE error box was covered. For those GRBs also seen by other satellites,
triangulation will reduce the error box size and may increase the
actual coverage.

\vspace{13pt}

\centerline{\it\bf Photographic Sky Patrol}

\vspace{8pt}

For the correlation of BATSE GRBs and photographic wide-field
plates a logistic network of 11 observatories with regular photographic
observations has been established, including Sonneberg, Tautenburg and
Hamburg (all FRG), Calar Alto (Spain), Ond\v{r}e\-jov (CR), Odessa,
Crimea (Ukraine), Dushanbe (Tadshikistan), Kiso (Japan),
Australian UK and ESO (La Silla, Chile). For nearly 60 GRBs detected by BATSE
since the launch of GRO simultaneous plates with typically
m$_{lim}\approx$2--3 mag for a 1 sec duration flash have been
identified$^{13,28,29}$.
No optical transient was found, resulting in limits for the {\bf flux} ratio
F$_\gamma$/F$_{opt}$ $\ge$ 1--20$^{13}$.

\vspace{13pt}
\centerline{\it\bf Neutrino and Muon Fluxes}
\vspace{8pt}

Searches for neutrinos, anti-neutrinos and muons during times of
BATSE GRBs have been performed using the data of the
Irvine-Michigan-Brookhaven (IMB) detector$^{21}$,
the Soudan 2 detector$^{20}$ and the Mont Blanc
Neutrino Telescope$^{22}$.
Both samples are believed to contain primarily atmospheric events,
i.e. neutrinos produced by the decay of secondary particles from cosmic-ray
interactions in the Earths atmosphere.
The IMB correlation with 183 Venera and Ginga GRBs
revealed no neutrino coincidences.
The resulting
flux limit of 1.7$\times$10$^{7}$ neutrinos cm$^{-2}$
corresponds to 2.7$\times$10$^{4}$ erg/cm$^2$ for neutrinos in excess
of 200 MeV$^{21}$.
Also, no muon coincidences were found.
A flux limit of 5.8$\times$10$^{-7}$ muons cm$^{-2}$
from neutrino interactions during GRBs is derived$^{21}$.

A correlation of 180 BATSE GRBs with the Soudan 2 muon detector events
revealed 18 trigger coincidences, near to the number expected by chance$^{20}$.
A correlation of the first 40 BATSE GRBs with events of the Mont Blanc
telescope revealed no excess interactions.
Depending on the neutrino flavour, the
upper limits range between 2.5$\times$10$^{10}$ cm$^{-2}$ s$^{-1}$
($\tilde{\nu}_e$ at 20$\le$E$_\nu\le$50 MeV) and
7.9$\times$10$^{12}$ cm$^{-2}$ s$^{-1}$
($\tilde{\nu}_{\mu+\tau}$ at 20$\le$E$_\nu\le$100 MeV)$^{22}$.

\vspace{13pt}

\centerline{\it\bf TeV Emission}

\vspace{8pt}

A correlation of the 1991/1992 season 0.4--4 TeV events observed with the
10m the Whipple Observatory Reflector with BATSE bursts has resulted in no
positive detection which allows to set limits on the density of primordial
black holes and the local density of cosmic strings$^{19}$.

The CYGNUS-I air shower array
detects ultra-high-energy (UHE) radiation at about 3.5 triggers/sec
for E$>$50 TeV from zenith and E$>$100 TeV for $\theta$$>$30\grad.
For a total of 52 (out of 260) GRBs
and 6 (out of 18) IPN GRBs having been in the CYGNUS FOV
($\theta$$<$60\grad) no evidence for UHE emission was found$^{16}$. The
limits for 3 GRBs are inconsistent with an extrapolation of the BATSE spectra,
indicating either a softening of the production spectrum at high energies
or the presence of UHE $\gamma$-ray absorption. If the latter possibility
is true this would imply cosmological GRB distances$^{16}$.

All GRBs of the
BATSE 2B catalog with a zenith angle $<$60\grad~ and an location error
$<$15\grad~ have been correlated with the HEGRA triggers. There are
simultaneous data for 85 bursts but no significant excess could be found
around the burst trigger times resulting in an upper limit between
10$^{-7}$...10$^{-10}$ cm$^{-2}$ s$^{-1}$ for energies in the range of
40--500 TeV$^{17}$.

\vspace{26pt}

\centerline{\bf NEAR-SIMULTANEOUS OBSERVATIONS}

\vspace{13pt}

\centerline{\it\bf Monitoring Projects}

\vspace{8pt}

Due to the nature of the monitoring observations it is natural that the
above discussed projects also allow searches for near-simultaneous
coverage of burst positions. While simultaneous photographic exposures$^{28}$
(12 hours before or after the burst) are available for dozens of GRBs with
typical limiting magnitudes of m$_{lim}$$\approx$7 mag for 1 sec flash
(and as short as about 1 hour after the burst), a recent
ETC observation of GB 941014 happened to be only 150 sec after the burst
onset. The 5 sec exposure with no optical transient sets a limit of
m$_{lim}$=10.2 mag$^{15}$.

A HEGRA search in a time window of 12 hours after the bursts also revealed
no significant excess rate$^{17}$. This excludes strong
extended TeV emission, as one could expect from the observation of the
extended GeV emission from GB 940217$^{30}$.

\vspace{13pt}

\centerline{\it\bf Rapid Follow-up Observations}

\vspace{8pt}

BACODINE$^{23}$ (BATSE Coordinates Distribution Network) automatically
evaluates the real-time telemetry stream from GRO, calculates approximate
coordinates for a BATSE burst and distributes this position to interested
observers. The COMPTEL/BATSE/NMSU network$^{24}$ derives improved positions
for those bursts occurring in the COMPTEL FOV using its imaging capabilities.
The characteristics of both these systems are given in the Table.

\small
\begin{table}[thbp]
\vspace*{-.2cm}
\begin{center}
\begin{tabular}{cccc}
\hline
\noalign{\smallskip}
               & {BACODINE} & {COMPTEL/BATSE/NMSU} & HETE \\
\noalign{\smallskip}
\hline
\noalign{\smallskip}
Time delay     & 5 sec            &   16 min            & 5 sec \\
Location error & $\pm$10\grad     &    $\pm$1-2\grad    & $\pm$10\amin \\
Rate           & 20--50 yr$^{-1}$ & 2--5 yr$^{-1}$ & 5--20 yr$^{-1}$ \\
\noalign{\smallskip}
\hline
\end{tabular}
\end{center}
\end{table}

\normalsize

{}From the wealth of observational results I want to pick out only three
examples, which should demonstrate the progress possible with the rapid
burst notification:

\noindent{\bf (1)}
The Gamma Ray Optical Counterpart Search Experiment (GROCSE) located
at Lawrence Livermore National Laboratory adapted a wide-FOV
telescope with a total of 23 cameras with 7\fdg7$\times$11\fdg5 FOV each
for optical burst follow-up observations$^{31}$. During
0.5 sec integration time a limiting magnitude of m$_{lim}$$\approx$8 mag
is reached, and images are taken at a repetition rate of every 5 sec.
During the first year of operation
8 BACODINE triggers
were received and observations of the burst locations started for two of
these events while the GRB was still bursting. For event 1 (2) the first
image was taken 24 (17) sec after the onset of the burst which lasted
40 (90) sec$^{32}$. While the analysis of these events still is in progress,
the limits achieved are considerably deeper than any previous observations.

\noindent{\bf (2)}
Within the observing campaign of GB 940301 by the BATSE/COMPTEL/ NMSU rapid
response network the COMPTEL error box was observed seven hours after
the burst with the 1m Schmidt telescope at Socorro reaching a limiting
magnitude of m$_{V}$ $\approx$ 16 mag$^{33}$. Still, no optical transient
was found. Radio observations at the same time at 8.42 GHz with the
34 m Goldstone antenna also found no new radio object$^{33}$.

\noindent{\bf (3)}
The bright gamma-ray burst GB 940301 was monitored at 1.4 GHz (2\fdg6 FOV)
and 0.4 GHz (8\fdg1 FOV) at the Dominion Radio Astrophysical Observatory (DRAO)
Synthesis Telescope starting starting 3 days after the burst$^{34}$.
A total of 245 radio sources were
identified in the summed image in an intensity range between 0.8--110 mJy,
but none was identified as a candidate for a flaring/fading radio counterpart
based on variability analyses.
The daily upper limits on the nondetection are $\approx$ 3.5 mJy
at 1.4 GHz and 55 mJy at 0.4 GHz and constrain some fireball parameters
for cosmological GRB models$^{34}$.

\vspace{26pt}

\centerline{\bf SUMMARY AND PROSPECTS}

\vspace{13pt}

Despite the increasing efforts in the search for GRB counterparts no
convincing candidate has been identified yet.
While the limits for simultaneous emission are still not yet deep enough,
those for up to a few hours after the bursts are getting constraining
for models which predict 1) predict strong and/or long-duration afterglows, or
2) simple extrapolations of the measured X-ray to $\gamma$-ray
spectra to lower and higher energies.

Observational improvements in the very near future can be expected mainly
from  (1) the
next generation device of GROCSE which will allow observations with a
smaller FOV down to m$_{lim}$$\approx$ 12--13 mag for a 1 sec flash$^{32}$, and
(2) the launch of HETE$^{35}$ which will measure burst positions with
better than 10\amin~ accuracy (see the Table) and after subsequent data
transmission to
ground observers will allow observations with smaller FOV telescopes
(and thus fainter limiting magnitudes) than feasible up to now.
This for the first time will allow optical observations with a sensitivity
which is comparable to the flux estimates deduced by extrapolating burst
spectra down from the X-ray/$\gamma$-ray range.

\vspace{26pt}

\centerline{\bf ACKNOWLEDGEMENTS}

\vspace{13pt}

I'm grateful to H.-S. Park, S. Barthelmy, T. Harrison and R.
Vanderspek for communicating their results prior to publication.
JG is supported by the Deutsche Agentur f\"ur Raum\-fahrt\-angelegen\-heiten
(DARA) GmbH under contract FKZ~50~OR~9201.

\vspace{13pt}

\centerline{\small\bf REFERENCES}

\vspace{8pt}
\small

\Ni ~1. Grindlay J.E., Wright E.L., McCrosky R.E., 1974. ApJ {\bf 192}, L113

\Ni ~2. Kasturirangan K., Rao, U., Sharma, D. \etal 1974. Nat. {\bf 252}, 113

\Ni ~3. Schaefer B.E., Bradt, H., Barat, C., \etal 1984. ApJ {\bf 286}, L1

\Ni ~4. Hudec R., Borovicka, J., Wenzel, \etal 1987. A\&A {\bf 175}, 71

\Ni ~5. Greiner J., Flohrer J., Wenzel W., Lehmann T., 1987. ApSS {\bf 138},
        155-171

\Ni ~6. Pizzichini G., Gottardi, M., Atteia, J., \etal 1986. ApJ {\bf 301}, 641

\Ni ~7. Bo\"er M., Hurley, K., Pizzichini, G., \etal 1991. A\&A {\bf 249}, 118

\Ni ~8. Greiner J.,  Bo\"er M., Motch C., \etal
     1991. 22nd ICRC Dublin, vol. {\bf 1}: 53-56

\Ni ~9. Greiner J., \etal
     1995. NATO ASI C450,
   eds. M.A. Alpar \etal, Kluwer, p. 519-522

\Ni 10. Bo\"er M., Greiner J., Kahabka P., \etal,
    1993. A\&A Suppl. {\bf 97}, 69

\Ni 11. Vrba F., Hartmann D.H., Jennings M.C., 1995. ApJ (in press)

\Ni 12. Sokolov V., \etal
       1995. Flares and Flashes, eds. J. Greiner \etal, LNP (in press)

\Ni 13. Greiner J., \etal
       1994. {\it Gamma-Ray Bursts}, eds. G.J. Fishman \etal,
        AIP 307: 408-412.

\Ni 14. Krimm H. \etal
     1994. {\it Gamma-Ray Bursts},
     eds. G.J. Fishman \etal, AIP 307: 423-427.

\Ni 15. Vanderspek R. \etal 1995. Flares and Flashes, eds. J. Greiner \etal,
       LNP (in press)

\Ni 16. Alexandreas D.E., Allen G.E., Berley D., \etal (CYGNUS Collab.),
       1994. ApJ {\bf 426}, L1

\Ni 17. Matheis V., 1995. Ph.D. thesis, Heidelberg University

\Ni 18. Aglietta M. \etal
     (EASTOP Collab.)
     1993. 23rd ICRC, Calgary 1993, vol. {\bf 1}: 61

\Ni 19. Connaughton V. \etal 1994. {\it Gamma-Ray Bursts}, eds.~G.J.~Fishman
       \etal,
       AIP 307:~470

\Ni 20. DeMuth D.M. \etal 1994. {\it Gamma-Ray Bursts}, eds. G.J. Fishman
\etal,
       AIP 307: 475

\Ni 21. LoSecco J.M., 1994. ApJ {\bf 425}, 217

\Ni 22. Aglietta M., Antonioli P., Badino G., \etal 1993, 23rd ICRC,
      Calgary 1993. vol. {\bf 1}: 69

\Ni 23. Barthelmy S. \etal
       1994.
       {\it Gamma-Ray Bursts}, eds. G.J. Fishman \etal, AIP 307: 643

\Ni 24. Kippen R.M., Ryan J.,  Connors A. \etal, 1995. this volume

\Ni 25. Schaefer B.E., 1994. {\it Gamma-Ray Bursts}, eds. G.J. Fishman \etal,
       AIP 307: 382-391.

\Ni 26. Hartmann D.H., 1995. {\it Flares and Flashes}, eds. J. Greiner \etal,
       LNP  (in press)

\Ni 27. Vanderspek R. \etal
       1994. {\it Gamma-Ray Bursts}, eds.~G.J.~Fishman~\etal, AIP~307: 438

\Ni 28. Greiner J., \etal 1995. {\it Flares and Flashes}, eds. J. Greiner
\etal,
       LNP (in press)

\Ni 29. Hudec R., 1995. {\it Flares and Flashes}, eds. J. Greiner \etal,
       LNP  (in press)

\Ni 30. Hurley K., Dingus B.L., Mukherjee R., \etal 1994. Nat. {\bf 372},
     652-654

\Ni 31. Akerlof C. \etal
     1994. {\it Gamma-Ray Bursts}, eds. G.J. Fishman \etal, AIP 307: 633-637.

\Ni 32. Park H.-S., 1994. AAS HEAD meeting, Napa Valley, Nov. 1994

\Ni 33. Harrison T.E., McNamara B.J., Pedersen H., \etal 1995. A\&A (in press)

\Ni 34. Frail D.A., Kulkarni S.R., Hurley K.C., \etal
  1995. ApJ {\bf 437}, L43-L46

\Ni 35. Ricker G. \etal 1988. {\it Nuclear Spectroscopy of Astrophysical
      Sources}, eds. N. Gehrels \hspace*{.7cm}\etal, AIP 170: 407

\end{document}